\documentclass[pra,showpacs,twocolumn,superscriptaddress,longbibliography]{revtex4-2}

\usepackage{amsmath}
\usepackage{amssymb}
\usepackage{graphicx}
\usepackage{dcolumn}
\usepackage{bm}
\usepackage{amsfonts}
\usepackage{subfigure}
\usepackage{float}
\usepackage{color}
\usepackage{xcolor}
\usepackage{braket}
\usepackage{hyperref}
\usepackage{comment}

\newcommand{\bv}[1]{\mathbf{ #1 }}

\newcommand{\fref}[1]{Fig.~\ref{#1}}

\newcommand{\eref}[1]{Eq.~(\ref{#1})}
\newcommand{\bref}[1]{(\ref{#1})}

\makeatletter
\def\maketitle{
\@author@finish
\title@column\titleblock@produce
\suppressfloats[t]}
\makeatother
\hypersetup{pdfborder = {0 0 0}}
\begin{document}

\title{Collapse of a quantum vortex in an attractive two-dimensional Bose gas}

\author{Sambit Banerjee}
\altaffiliation{These authors contributed equally to this work.}
\affiliation{Department of Physics and Astronomy, Purdue University, West Lafayette, IN 47907, USA}
\author{Kai Zhou}
\altaffiliation{These authors contributed equally to this work.}
\affiliation{Department of Physics and Astronomy, Purdue University, West Lafayette, IN 47907, USA}

\author{Shiva Kant Tiwari}
\altaffiliation{These authors contributed equally to this work.}
\affiliation{Department of Chemistry, Purdue University, West Lafayette, IN 47907, USA}
\author{Hikaru Tamura}
\altaffiliation{Present address: Insitute for Molecular Science, Okazaki, Aichi 444-8585, Japan}
\affiliation{Department of Physics and Astronomy, Purdue University, West Lafayette, IN 47907, USA}
\author{Rongjie Li}
\affiliation{Department of Physics and Astronomy, Purdue University, West Lafayette, IN 47907, USA}
\author{Panayotis Kevrekidis}
\affiliation{Department of Mathematics and Statistics, University of Massachusetts Amherst, Amherst, Massachusetts 01003-9305, USA}
\author{Simeon I. Mistakidis}
\affiliation{Department of Physics, Missouri University of Science and Technology, Rolla, MO 65409, USA}
\author{Valentin Walther}
\affiliation{Department of Physics and Astronomy, Purdue University, West Lafayette, IN 47907, USA}
\affiliation{Department of Chemistry, Purdue University, West Lafayette, IN 47907, USA}
\affiliation{Purdue Quantum Science and Engineering Institute, Purdue University, West Lafayette, IN 47907, USA}
\author{Chen-Lung Hung}
\email{clhung@purdue.edu}
\affiliation{Department of Physics and Astronomy, Purdue University, West Lafayette, IN 47907, USA}
\affiliation{Purdue Quantum Science and Engineering Institute, Purdue University, West Lafayette, IN 47907, USA}

\date{\today }

\begin{abstract}
We experimentally and numerically study the collapse dynamics of a quantum vortex in a two-dimensional atomic superfluid following a fast interaction ramp from repulsion to attraction. We find the conditions and time scales for a superfluid vortex to radially converge into a quasi-stationary density profile, demonstrating the spontaneous formation of a vortex soliton-like structure in an atomic Bose gas. We record an emergent self-similar dynamics caused by an azimuthal modulational instability, which amplifies initial density perturbations and leads to the eventual splitting of a solitonic ring profile or direct fragmentation of a superfluid into disordered, but roughly circular arrays of Townes soliton-like wavepackets. These dynamics are qualitatively reproduced by simulations based on the Gross-Pitaevskii equation. However, a discrepancy in the magnitude of amplified density fluctuations predicted by our mean-field analysis suggests the presence of effects beyond the mean-field approximation. Our study sets the stage for exploring out-of-equilibrium dynamics of vortex quantum matter quenched to attractive interactions and their universal characteristics.
\end{abstract}

\maketitle

Vortices are prevalent fundamental excitations in nonlinear fields~\cite{pismen1999vortices}. Probing vortex dynamics has played a pivotal role in studies of condensed matter, quantum gases, and nonlinear optics, from developing better understandings of superconductivity and superfluidity~\cite{salomaa1987quantized,blatter1994vortices,zwierlein2005vortices} to finding new applications using angular momentum carrying optical beams~\cite{shen2019optical}. Generally, the stability of vortices relies on the nature of the nonlinearity involved.  
As vortices are described by a field with integer multiples of $2\pi$-phase winding around a phase singularity, which necessitates a zero amplitude defect at the vortex core, a self-defocusing (repulsive) nonlinearity can smoothen and stabilize the wave away from the defect. With a self-focusing (attractive) interaction, on the other hand, such waveforms become unstable against wave collapse. Nevertheless, it has been shown to be possible to embed vorticity in a stationary state, called a vortex soliton~\cite{kruglov1992theory,desyatnikov2005optical,malomed2019vortex}, where the self-focusing effect balances the wave dispersion of a ring-shaped waveform with phase winding. 

However, even without radial wave collapse, a vortex soliton is still unstable against a pattern-forming instability~\cite{firth1997optical}, with which any azimuthal wave perturbations beyond a critical length scale can self-amplify. This instability leads to the growth of modulations in discrete angular modes and can eventually fragment a ring-shaped vortex soliton into a circular array of solitary waves in a distinct angular pattern like a `necklace'~\cite{firth1997optical,petrov1998observation,bigelow2004breakup}. 
To-date, experimental studies of solitons with vorticity focus on engineered optical vortices in nonlinear media~\cite{petrov1998observation,desyatnikov2005optical, malomed2019vortex,swartzlander1992optical,duree1995dark,chen1997steady,soljavcic1998self,jeng2004partially,reyna2016robust,reyna2020observation,bigelow2004breakup,vuong2006collapse} and photonic lattices~\cite{fleischer2004observation,neshev2004observation,bartal2005observation}.

In quantum gas experiments, to our knowledge, vortex dynamics with attractive interactions has remained completely unexplored. Despite existing discussions on the formation and stability of vortex solitons in Bose-Einstein condensates~\cite{saito2002split, carr2006vortices, mihalache2006vortex}, a successful demonstration of a collapsing vortex has remained elusive. This is hindered by state preparation in attractive Bose gases that have strong tendencies to collapse~\cite{sackett1998growth,gerton2000direct,donley2001dynamics,roberts2001controlled,eigen2016observation}. 
Whether a many-body vortex soliton can form spontaneously also remains an open question. 

Recently, optical box-confined quantum gases have emerged as excellent platforms~\cite{navon2021quantum} for studying wave collapse~\cite{eigen2016observation,chen2020observation} and pattern-forming instabilities~\cite{zhang2020pattern,chen2020observation,tamura2023observation,hernandez2024connecting,liebster2025observation}, due to the ability to remove undesired trap length scales and the flexibility to adjust the box boundary conditions. It thus becomes possible to access intrinsically unstable many-body states through nonequilibrium quench dynamics. By using a magnetic Feshbach resonance to quench the atomic interaction from repulsion to attraction, self-trapped, yet unstable fundamental solitons in two dimensions (2D)--the Townes solitons~\cite{chiao1964self}--were found, surprisingly, following a universal wave collapse dynamics of a modulational instability (MI) in 2D~\cite{chen2020observation,chen2021observation}. 

\begin{figure}[b]
\includegraphics[width=\columnwidth]{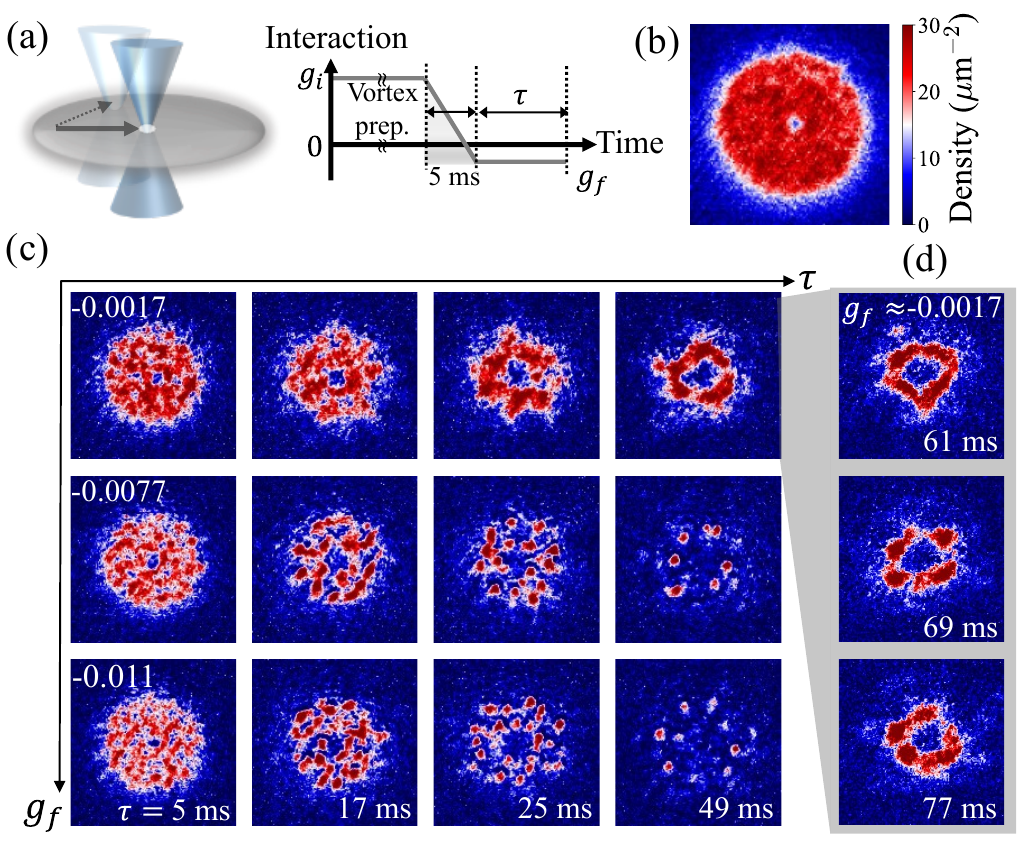}
\caption{ Creation and collapse of a quantum vortex. (a) Experimental scheme. A single vortex is created at the center of a circular 2D superfluid, followed by simultaneous ramp-down of the in-plane confinement and the interaction parameter to $g_f<0$ for a TOF time $\tau$ in the 2D plane. (b) Density image taken after the ramp at $\tau=5$~ms, $g_f\approx -0.0077$, and averaged over 4 shots. (c, d) Single-shot images with variable interaction $g_f$ and time $\tau$ (c) and $g_f\approx -0.0017$ taken at longer times (d). Ring-like and necklace-like solitonic structures are observed at sufficiently long TOF times. Image size: (b) $(56\mu$m$)^2$ and (c, d) $(65\mu$m$)^2$. All images use the same color scale.}
\end{figure}
 
Watching vortices collapse following an interaction quench in a quantum gas can open pathways to uncover facinating self-patterned structures or vortex solitons. Collapse dynamics of vortices can exhibit new self-similar scaling behaviors and time scales distinct from those of wave collapse in vortex-free, non-rotating Bose gases~\cite{chen2020observation}. Azimuthal MI in a quantum vortex, when seeded by zero-temperature quantum fluctuations, can serve as a `quantum parametric amplifier'~\cite{esteve2008squeezing,chen2021observation2}. This may lead to macroscopic quantum entanglement and even many-body fragmentation~\cite{nozieres1982particle,wilkin1998attractive,spekkens1999spatial,ho2000fragmented,mueller2006fragmentation,nguyen2019parametric,evrard2021observation} in the angular momentum states that can have applications in quantum metrology~\cite{luo2022quantum,pezze2018quantum}. 

\begin{figure*}[t!]
\includegraphics[width=0.8\textwidth]{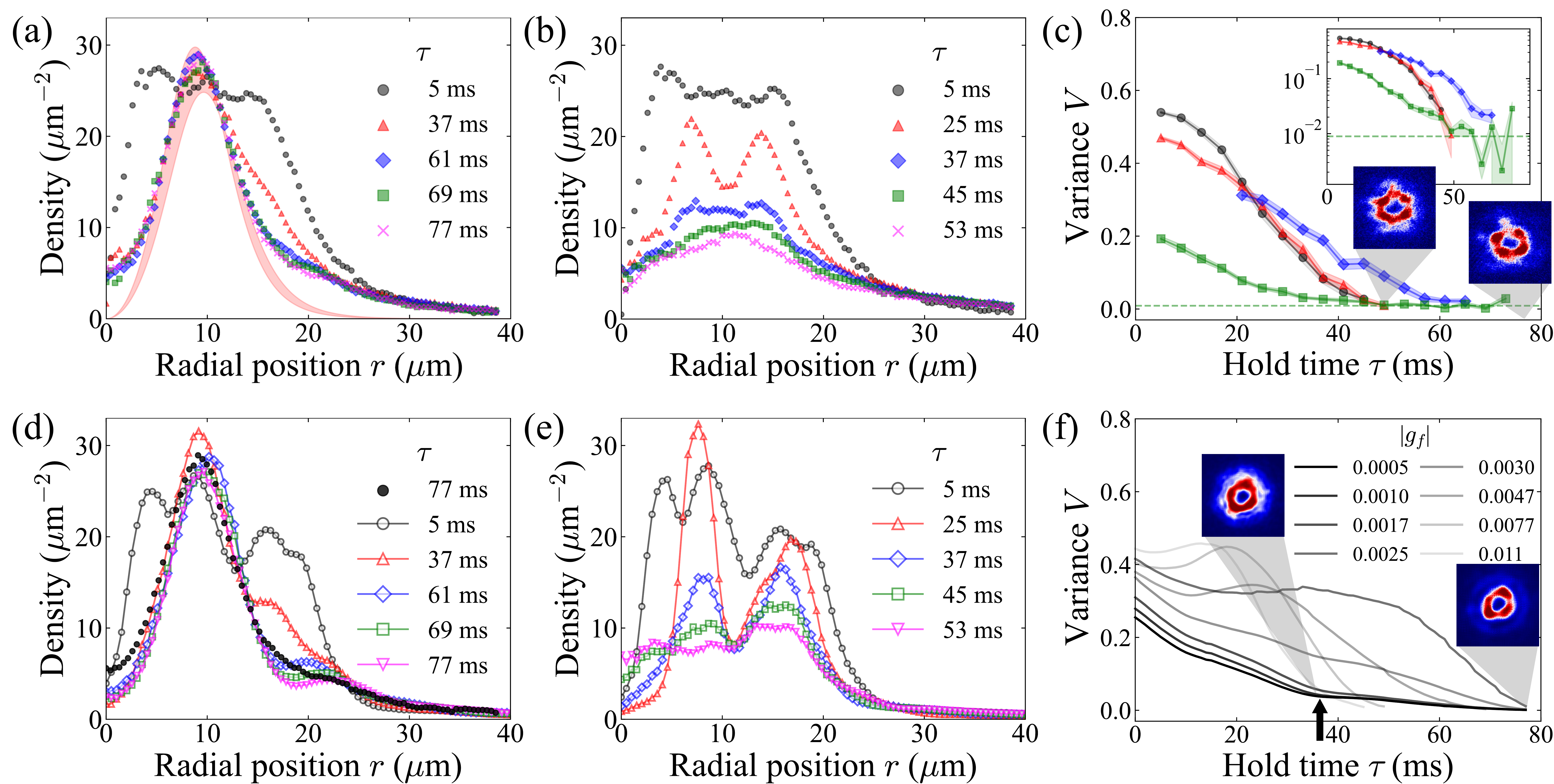}
\caption{\label{fig2_radial} Evolution of the radial density profiles. Experiment: (a, b) Radially averaged density profiles (filled symbols) for (a) $g_f\approx-0.0017$ and (b) $g_f\approx-0.0077$ measured at the indicated time $\tau$. Shaded band in (a) shows the stationary vortex soliton density profile rescaled to match the peak densities measured at $\tau=37\sim77$~ms. 
(c) Variance $V$ of the measured density profiles relative to a reference profile recorded at time $\tau_\mathrm{ref}$ for $g_{f}\approx -0.011$ (black circles), $-0.0077$ (red triangles), $-0.0047$ (blue diamonds), and $-0.0017$ (green squares). Shaded bands indicate the statistical errors. Inset shows $V$ in logarithmic scale and averaged density images at $\tau=49$~ms and 77~ms for $g_f \approx -0.0017$, respectively. Dashed line indicates the averaged value of $V$ at $g_{f}\approx -0.0017$ between $49\sim77$~ms. Simulation: (d, e) Radial density profiles evaluated from the GPE simulations (open symbols) using experimental parameters as in (a) and (b), respectively. Filled circles illustrate a density profile from (a) at time $\tau>\tau_r$. (f) Variance $V$ of simulated density profiles. Arrow indicates the onset of a plateau. The insets present the corresponding averaged density images for $g_f \approx -0.0017$ at the indicated times.}
\end{figure*}

In this letter, we study the collapse dynamics of a single vortex in attractive Bose gases confined in a quasi-2D box, and report the spontaneous formation of vortex soliton-like ring structures and self-patterned Townes soliton necklaces. We create a single vortex nearly deterministically in a circular 2D superfluid, followed by ramping the atomic interaction to the weakly attractive regime and performing 2D time-of-flight (TOF) imaging. We observe clear radial convergence towards a vortex soliton density profile and record a new self-similar scaling behavior and time scale of an azimuthal MI that either splits a vortex soliton-like ring profile or directly fragments a superfluid into disordered, but roughly circular arrays of Townes soliton-like wavepackets. These dynamics are qualitatively reproduced by the 2D Gross-Pitaevskii equation (GPE) incorporating a three-body loss term and initial random noise simulating zero-temperature phonon fluctuations. We found discrepancies in the magnitude of amplified density fluctuations between experiment and simulations, which hint at beyond mean-field effects in the fragmentation dynamics. Nevertheless, our results provide unprecedented insights into the self-similar collapse dynamics of a quantum vortex. 

The starting point of the experiment is a homogeneous 2D Bose gas of $3.8\times 10^4$ cesium atoms trapped in a circular box and prepared deeply in the superfluid regime~\cite{supp}. The initial radius and density of the 2D superfluid is $r_i\approx 21~\mu$m and $n_i\approx25~\mu$m$^{-2}$, respectively. The interaction strength $g$ is controlled by a magnetic Feshbach resonance, setting the initial value to $g_i\approx0.1$. We create a single vortex at the center of the 2D superfluid using a `chopstick' method demonstrated in Refs.~\cite{samson2016deterministic,gertjerenken2016generating,kwon2021sound} and illustrated in Fig.~1(a) and (b).  

To induce collapse dynamics following the vortex preparation, we ramp the interaction strength in 5~ms to a variable attractive value $g_f$, as schematically depicted in Fig.~1(a), while simultaneously ramping off the in-plane box confinement (in 3~ms). The ramp speed is fast compared to subsequent vortex dynamics, but slow enough to avoid creating additional excitations near the edge of the 2D gas, such as ring dark solitons~\cite{tamura2023observation}. We then allow the gas to freely evolve in the 2D plane for a variable TOF time $\tau$, followed by absorption imaging to record the density distribution.

Figure 1(c) shows single-shot images of samples held at different interaction strengths. For weak attraction $g_f\approx-0.0017$ (top row), the vortex core soon expands to a larger size and the disk-like density structure evolves into a ring-shaped profile. At longer hold times, the ring structure becomes distorted and splits, manifesting the presence of the azimuthal MI~\cite{caplan2009azimuthal}; see Fig.~1(d). At more negative values of $g_{f}$, the 2D gas directly fragments into circular arrays of density blobs, as shown in the second and the third rows of Fig.~1(c). This peculiar multi-ring fragmentation indicates fast wave collapse, both radially and azimuthally. 
One expects the blobs to form around the size of an interaction length $\xi = \pi/k_\mathrm{MI}$, where $k_\mathrm{MI}=\sqrt{2 n_i |g_f|}$ is the most unstable wavenumber with the largest imaginary Bogoliubov dispersion~\cite{chen2020observation}.
At longer TOF times, the circular arrays appear to collide with each other in the radial direction and collapse, leaving behind a disordered array of blobs whose density profiles approach those of Townes solitons (see \cite{supp}). We dub this disordered array a Townes soliton necklace. 

The multi-ring collapse dynamics can be analyzed through radially averaged density profiles over $\sim 12$ samples, where we set the vortex core as the origin (Fig.~2). At $g_f\approx-0.0077$ (Fig.~2(b)), two peaks initially appear, corresponding to the formation of two approximately circular arrays of blobs as shown in Fig.~1(c). These peaks soon collapse, forming a broad peak contributed from the radial average of a Townes soliton necklace and the collision remnant. 

Remarkably, the radial profile of the ring-shaped structure observed at the weaker interaction $g_f\approx -0.0017$ remains nearly stationary for more than 40~ms from $\tau \gtrsim 37~$ms; see Fig.~2(a). This suggests that the vortex evolves towards a self-trapped vortex soliton. The peak density is found at a fixed radial position at $r\approx 10~\mu$m, agreeing fairly well with a stationary-state solution of the GPE~\cite{carr2006vortices}, $n(r)=n_p\left|\phi_{\mathrm{vs}}\left(\sqrt{n_p\left|g_f\right|} r\right)\right|^2$, where $n_p$ is the peak density and $\phi_{\mathrm{vs}}$ is the radial part of an ideal vortex wavefunction $\psi(R,\theta)=\phi_\mathrm{vs}(R)e^{i s\theta}$ with $s=\pm 1$. We solve the radial wavefunction using the 2D GPE in a scale-invariant form $-\frac{1}{2}\left(\frac{\partial^2 }{\partial R^2}+\frac{1}{R} \frac{\partial }{\partial R} -\frac{s^{2} }{R^2}\right)\phi_\mathrm{vs}-|\phi_\mathrm{vs}|^2 \phi_\mathrm{vs}=\tilde{\mu} \phi_\mathrm{vs}$, where $R=\sqrt{n_p\left|g_f\right|}r$ is the rescaled radial coordinate and $\tilde{\mu}$ is the chemical potential. 
We note this is a unique, scale-invariant vortex solution in 2D, with $n_p$ fixing the physical size of the wavefunction, and no free parameters are used in the comparison. The measured quasi-stationary radial density profiles nevertheless deviate from the ideal solution near the low-density wings, potentially due to very slow collapse dynamics of atoms dispersed during the ramp-down of the box confinement.

\begin{figure}[!t]
\includegraphics[width=1\columnwidth]{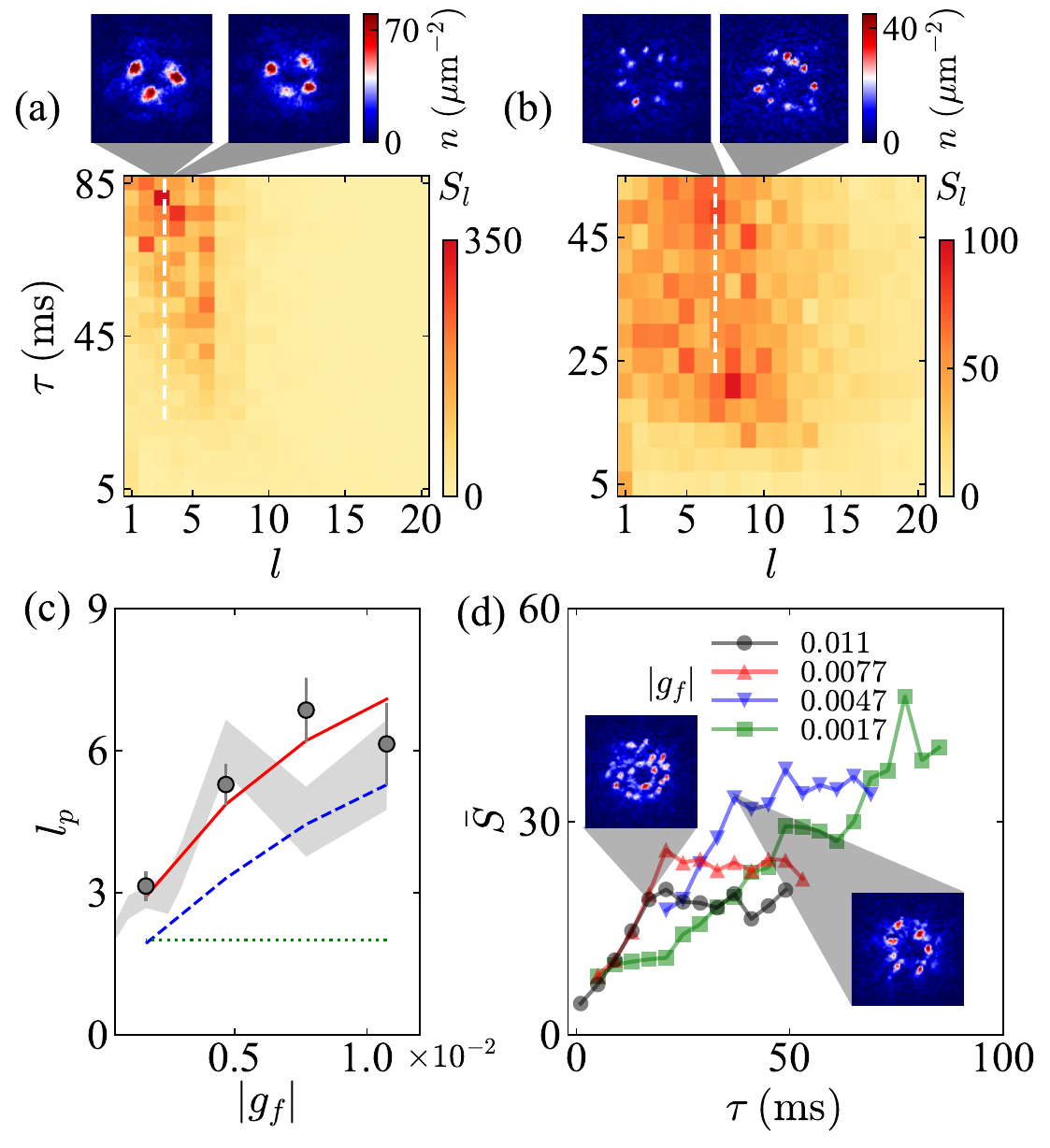}\label{fig3}
\caption{Dynamics of azimuthal modulational instability. (a) The angular power spectrum $S_l$ at $g_f\approx -0.0017$ for different TOF times $\tau$. Single-shot images at the top are recorded at $\tau=77~\mathrm{ms}$, representing soliton splitting into an  angular mode at $l=3$ and 4, respectively.  (b) $S_l$ at $g_f\approx -0.0077$. Single-shot images are recorded at $\tau=53~\mathrm{ms}$, evincing dominant mode(s) $l=7$ and $9\sim11$, respectively. Vertical dashed lines in (a) and (b) indicate the peak position $l_{\mathrm{p}}$~\cite{supp}. (c) $l_p$ versus $g_f$ (filled circles). Error bars represent the standard error of the mean. Blue dashed line marks $l_p = \bar{r}k_\mathrm{MI}$. Red solid line is a fit.  
Green dotted line marks the most unstable mode of a vortex soliton. Shaded band shows $l_p$ and the standard error evaluated from the GPE simulations. (d) Time evolution of the averaged power spectrum $\bar{S}$ (color symbols).  
Insets present single-shot density images taken at the saturation time of $\bar{S}$, $\tau_\theta=17~$ and 37~ms, for $g_f\approx-0.011$ and $-0.0047$, respectively. 
}
\end{figure}

We can quantify when the radial quasi-stationarity is reached, to within experimental noise, by comparing the radial density profiles $n(r,\tau)$ to a reference profile obtained at a long TOF time $\tau_{\mathrm{ref}}$, 
chosen to be 4~ms after the last point of each curve in Fig.~2(c). We compute a time-dependent variance $V(\tau)\equiv \int [n(r,\tau) - n(r,\tau_{\mathrm{ref}})]^2 dr/\int n(r,\tau)^2 dr$. If the density radially converges to a quasi-stationary profile, $V$ reduces to a small value until it becomes limited by technical noise or residual radial dynamics, and plateaus at some intermediate time $\tau<\tau_{\mathrm{ref}}$. Indeed, the density profiles of $g_f \approx -0.0017$ satisfy this criterion with $V$ plateauing at 1\% level of the integrated density profile after time $\tau_r \approx 49$~ms. For more attractive samples, $V$ decreases monotonically, implying the absence of a radially quasi-stationary intermediate state. 

We attempt to reproduce key signatures of the radial collapse dynamics in the time evolutions of the 2D GPE \cite{supp}, see Figs.~\ref{fig2_radial}(d-f). To model atom losses during the collapse dynamics, a three-body loss term is incorporated in the GPE simulation (see~\cite{supp}). 
The density profiles in Fig.~\ref{fig2_radial}(d) match well with the experimental data in Fig.~\ref{fig2_radial}(a) and converge at the same time scale. In Fig.~\ref{fig2_radial}(f), we find that the simulated variance $V$ indeed converges to $\lesssim3.5\%$ of the integrated density profile for $|g_f|\lesssim 0.0025$ at $\tau \gtrsim 40~$ms, and slowly descends due to residual evolution in the low-density wing \cite{plateau} and accumulated three-body losses, as seen in Fig.~\ref{fig2_radial}(d). 

The insets in Fig.~2(c) and (f) compare experimental and numerical density images (each averaged over $\sim 10$ shots) measured at $\tau=\tau_r$, when the radial density profiles converge, and near the end of our observation window at $\tau=77$~ms. We note that, while the vortex soliton-like profile has remained nearly unchanged in the averaged density, the ring structure in single shots already suffers significant azimuthal distortions (Fig.~1(d)), which we attribute to the presence of an azimuthal MI. 

We now ask how and when the azimuthal instability manifests. To analyze the dynamics, we evaluate the azimuthal number distribution $N_{\theta}=\int^{r_i}_0 \ n(r,\theta)rdr$ and calculate the angular power spectrum $S_l=\langle |N_l|^2\rangle/N$, where $l\in \mathbb{N}$ labels discrete angular modes, $N_l=\int^{2\pi}_0 N_{\theta}e^{-il\theta} d\theta$, $N$ is the total atom number, and $\langle \cdots\rangle$ denotes ensemble averaging. 

In Figs.~3(a) and (b), we present the time evolution of measured spectra $S_l$, revealing dynamic competition within a band of unstable angular modes. Each spectrum shows significant growth initially at some larger angular modes near $l=6$ and 11, respectively, which is succeeded by a band of lower $l$ modes that ultimately culminate in a distinct peak. The peak is broader for more attractive interactions, as in Fig.~3(b), where smaller-scale perturbations also become unstable. The prevailing peak in the power spectrum reveals the most unstable $l$-mode(s) that could most frequently fragment a ring structure into $l$ pieces--a signature of pattern-forming instability dynamics. In the top panels of Figs. 3(a) and (b), we show single-shot density images to illustrate the ensuing fragmentation caused by different modes.

In Fig.~3(c), we extract the peak position $l_p$ measured at later times in the spectra \cite{supp} and compare it with theoretical expectations. Overall, the peak locations follow reasonably well the prediction from a variational calculation $l_p \approx \bar{r} k_\mathrm{MI}=\pi\bar{r}/\xi$ ~\cite{caplan2009azimuthal}. Here the peak azimuthal mode number is determined by the most unstable wavenumber $k_\mathrm{MI}$ and an effective radius $\bar{r} \approx 7.5~\mu$m, evaluated using the density profiles near the fragmentation time \cite{supp}. We also fit $l_p$ using a variable $\bar{r}$. The fitted $\bar{r}\approx 10~\mu$m appears to reflect the peak radii of actual samples as shown in Fig.~2. Meanwhile, we find qualitatively similar power spectra from the GPE data and very good agreement with the measured $l_p$. The overall trend reflects that the interaction length $\xi\propto k_\mathrm{MI}^{-1}$ is a dominant length scale fragmenting our atomic samples. 
For an ideal vortex soliton wavefunction, on the other hand, 
the expected most unstable mode $l=2$ is independent of the values of $(g_f, n_p)$~\cite{supp,saito2002split,mihalache2006vortex}.

\begin{figure}[t!]
\includegraphics[width=1\columnwidth]{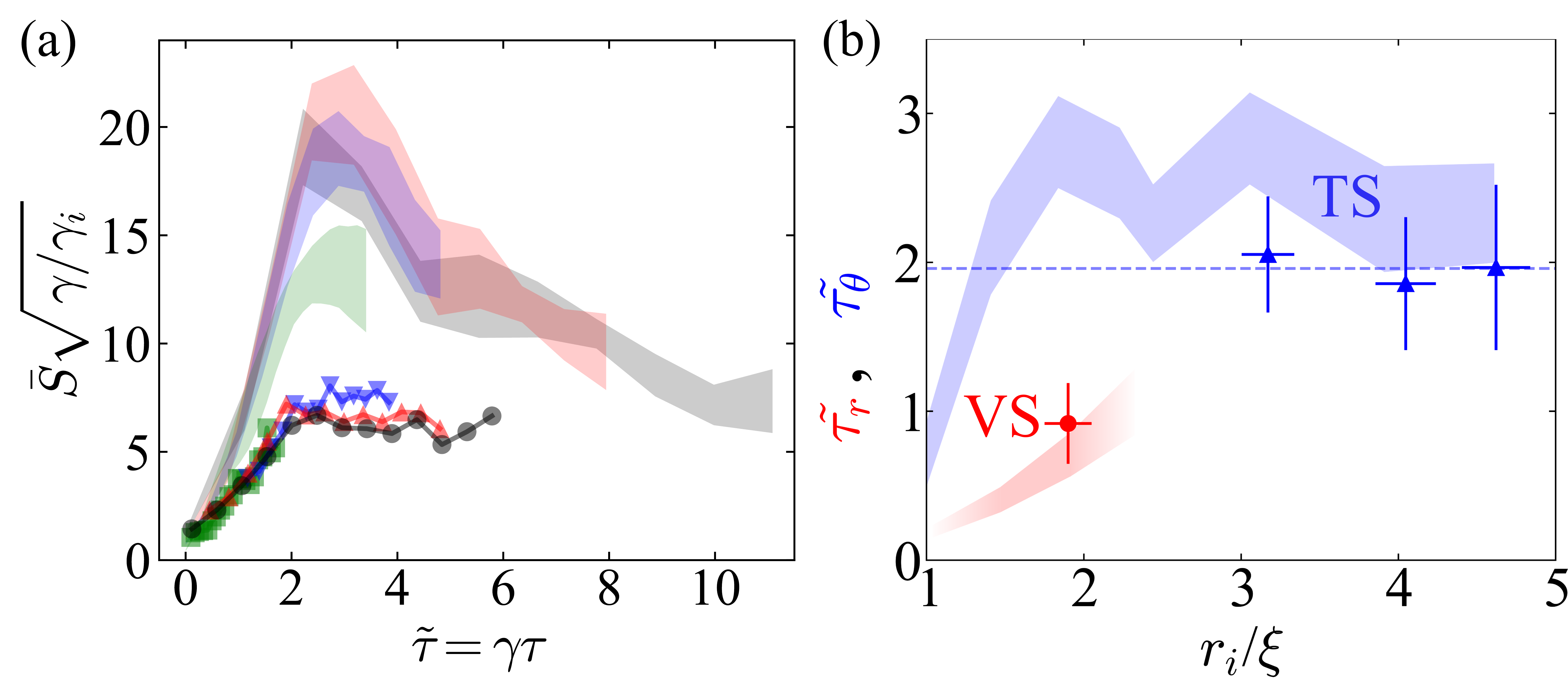}\label{fig4}
\caption{Self-similar scaling behavior and collapse timescales.  
(a) Scaled power spectra (color symbols, same as Fig.~3(d)) and simulations (shaded bands, same color) versus rescaled time. (b) Collapse time scales versus rescaled initial sample size. Formation of vortex soliton-like ring structures (VS) at $\tilde{\tau}_r=\gamma \tau_r\lesssim 1$ is observed for a sample with $r_i/\xi\lesssim 2$ (red circle). Samples with $r_i/\xi \gtrsim 3$ feature direct fragmentation to Townes solitons (TS) at nearly identical times $\tilde{\tau}_\theta=\gamma\tau_\theta\approx 2$ (blue triangles). Blue dashed line marks the mean value. Shaded bands indicate the time and uncertainty for radial convergence (red) and azimuthal fragmentation (blue) determined from simulations.
}
\end{figure}

Surprisingly, we find a self-similar scaling behavior and a universal time scale when the azimuthal MI fragments a sample. To see this, we first calculate $\bar{S}$, the mean power spectrum averaged over $l=1\sim 30$, and monitor its time evolution in Fig. 3(d). For every interaction except $g_f\approx -0.0017$, the growth of $\bar{S}$ saturates, indicating when the azimuthal perturbations cease to grow and the samples fragment. The insets are representative single-shot density images taken at the onset of saturation ($\tau_\theta$). In Fig. 4(a), the scaled spectra $(\gamma/\gamma_i)^\alpha\bar{S}$ versus the rescaled time $\tilde{\tau}=\gamma\tau$ clearly exhibit a self-similar behavior, particularly in the early-time growth dynamics, where $\alpha=0.5$ is an empirical scaling exponent and $\hbar \gamma_{(i)}=\hbar^2 n_i|g_{f(i)}|/m$ is the (initial) interaction energy.  
This behavior suggests that, regardless of the radial dynamics, azimuthal perturbations exhibit a self-similar growth curve with a peculiar interaction scaling and fragmentation occurs at a nearly fixed time $\tilde{\tau}_\theta\approx 2$ despite manifesting by different angular patterns.  
Rescaling the corresponding GPE data, we find a consistent self-similar scaling behavior with a peak at $\tilde{\tau}_\theta\approx 2.5$. However, nontrivial discrepancies exist between experiments and simulations in the overall magnitude of the rescaled spectra before fragmentation and in the exact quantitative features after collapse, which cannot be resolved even after extensively considering systematic effects within our mean-field analyses; see~\cite{supp} for details. Lastly, we note that a scaling behavior of $\alpha=1$, predicted by a linear stability analysis of the MI, and wave fragmentation at $\tilde{\tau}\approx 0.8$ was observed in vortex-free 2D gases~\cite{chen2020observation}. The reduced scaling exponent $\alpha=0.5$ is empirical and still awaits a theoretical explanation. Moreover, our observation suggests that the presence of a single vortex significantly delays azimuthal wave fragmentation.

In Fig.~4(b), we summarize the fate of a vortex following a fast ramp to attractive interactions. Based on the scale-invariant nature of weakly interacting 2D Bose gases~\cite{hung2011observation,yefsah2011exploring,chen2021observation,bakkali2021realization}, we expect that the observed dynamics can be replicated in samples of different initial sizes and final interaction strengths within the same range $1 \lesssim r_i/\xi \lesssim 5$, provided that the weak three-body loss dynamics does not significantly break the scale-invariance. For samples with $1<r_i/\xi \lesssim 2$, we observe the formation of a vortex soliton-like ring profile at $\tilde{\tau}=\tilde{\tau}_r \lesssim 1$. Perturbations in the ring profile grow continuously until approaching $\tilde{\tau} \approx 2$ where the density ring nearly splits into a circular array of bright solitons with distinct angular patterns (Fig.~3(a)). Soliton splitting speeds up when the interaction length approaches the sample initial size $\xi\approx r_i$, as confirmed by GPE simulations \cite{supp}. On the other hand, for larger size or more attractive samples with $r_i/\xi \gtrsim 3 $, more than one density ring forms in the radial direction. These samples all fragment azimuthally at $\tilde{\tau}=\tilde{\tau}_\theta\approx 2$. The multi-ring structure eventually collapses, creating a disordered Townes soliton necklace. We note that starting with different initial interaction energy $\hbar \gamma_i$ and temperature $T$ can change the amplitudes of initial density fluctuations that seed the instability. Our samples have suppressed thermal fluctuations due to $\eta = \hbar\gamma_i/k_B T\gtrsim 1$. When increasing $\eta>1$, we expect $\tilde{\tau}_\theta, \tilde{\tau}_r$ to remain roughly unchanged, as the instability remains primarily seeded by quantum fluctuations. For samples with lower $\eta<1$, however, increased initial thermal fluctuations and density perturbations may lead to early fragmentation, making a solitonic ring profile shorter-lived with smaller $\tilde{\tau}_\theta< 2$.

Our observed quench dynamics point towards a universal behavior and extended time scales to create self-trapped vortex quantum matter. The measured mean radial density profiles are well captured by the mean-field model. However, the discrepancy in the amplitude of the azimuthal power spectra, which reflect the density fluctuations and possibly emergent correlations, may suggest the presence of beyond mean-field effects not included in our GPE analyses. This calls for further theoretical investigations, for instance, ranging from perturbative techniques~\cite{petrov2024beyond} to sophisticated ab-initio approaches~\cite{cao2017unified}. A similar dynamics can also be investigated in two-component Bose gases with an effective attractive interaction  \cite{bakkali2021realization,romero2024experimental}. Further manipulation or studies of the azimuthal MI of a vortex soliton can be carried out by adding a ring trap for radial confinement or by pinning the vortex core with a repulsive beam. Another nontrivial extension would be to study the quench dynamics of a vortex lattice in a superfluid~\cite{madison2001stationary,abo2001observation}. By tuning the vortex density to reach a lattice constant $\gtrsim 2\xi$, roughly corresponding to $r_i/\xi\gtrsim 1$ for the size of individual vortex puddles in the lattice, one may find novel vortex soliton-like arrays and can study the subsequent collision and collapse of these solitonic structures. While our work is based on weakly attractive Bose gases, our experimental method can be adapted to embed vortices in a quantum droplet~\cite{ferrier2016observation,chomaz2016quantum,semeghini2018self, cabrera2018quantum}, where wave collapse from a strong mean-field attraction is prevented by an effective repulsion from quantum fluctuations~\cite{bulgac2002dilute,petrov2015quantum,naidon2021mixed}, and a self-trapped state with vorticity will exhibit different instability behaviors~\cite{li2018two,kartashov2018three,dong2021rotating}.

\bibliography{library}

\begin{thebibliography}{81}
\bibitem [{pla()}]{plateau}%
  \BibitemOpen
  \href@noop {} {}\bibinfo {note} {When evaluating $V$ for $0<r\leq14~\mu$m in order to exclude the low-density wings, we find that $V$ plateaus at $0.7\%$ ($<2\%$) for experiment (GPE).}\BibitemShut {Stop}%
\end{thebibliography}

\section*{Acknowledgments}
We thank Qiyu Liang, Cheng Chin, Moti Segev, and Ron Ziv for fruitful discussions. This work is supported in part by the NSF (Grant \# PHY-1848316 \& PHY-2409591), the AFOSR (FA9550-22-1-0327), and the DOE QuantISED program through the Fermilab Quantum Consortium. P.G.K. acknowledges support by the NSF under the awards PHY-2110030 and DMS-2204702. R. Li acknowledges the Rolf Scharenberg Graduate Summer Research Fellowship.

\appendix
\renewcommand{\thesection}{}
\renewcommand{\thesubsection}{S\arabic{subsection}}
\renewcommand{\figurename}{Fig.}
\renewcommand{\thefigure}{SM\arabic{figure}}
\setcounter{figure}{0}
\renewcommand{\theequation}{S\arabic{equation}}
\setcounter{equation}{0}
\setcounter{page}{1}

\title{Supplemental Material for:\\Collapse of A Quantum Vortex in An Attractive Two-Dimensional Bose Gas}
\maketitle
\onecolumngrid
\subsection{The box potential}\label{App:box}
The two-dimensional (2D) homogeneous Bose gases are prepared in a flat optical box potential. The in-plane confinement is provided by a repulsive ring beam, created by a blue-detuned 780 nm light reflecting off a digital mirror device (DMD) and projected through a microscope objective (numerical aperture $\mathrm{N.A.}=0.6$). The same DMD is also used to dynamically project additional repulsive potentials to create a single vortex at the center of the 2D superfluid. 
The out-of-plane (vertical) confinement of the 2D gas is provided by a single node of a repulsive optical lattice potential, formed using another 780 nm light. The out-of-plane atomic motion is frozen in the ground state with an oscillator length $l_{z}\approx 265$~nm. The interaction strength $g=\sqrt{8\pi}a_{s}/l_{z}$ 
is controlled by a magnetic Feshbach resonance that tunes the $s$-wave scattering length $a_{s}$. The 2D gas lies deeply within the superfluid regime with an initial temperature $T< 8~$nK, much lower than the critical temperature for the Berezinskii-Kosterlitz–Thouless superfluid phase transition $T_{c} \approx \frac{2\pi\hbar^2n_i}{mk_B\ln(380/g_i)} \approx 54$~nK~\cite{prokof2001critical,hung2011observation}. Here, $k_B$ is the Boltzmann constant, $g_i\approx0.1$ the initial interaction strength, $n_i\approx25~\mu$m$^{-2}$ the initial density, $\hbar=h/2\pi$ the reduced Planck constant, and $m$ the atomic mass.

We note that, without compensation, a very weak in-plane trap corrugation is present in the box due to the confining lattice beam. The potential variation is estimated to be $\Delta U \lesssim k_B\times 2~$nK. While it imposes no visible effect on the initial 2D superfluid, a weakly attractive sample could eventually be localized in the local potential minima created by this corrugation, forming linear density stripes. To mitigate this effect, we ramp on a compensating potential pattern at $\tau=2~\mathrm{ms}$ using the DMD with a matching periodicity ($\approx 18~\mu$m) to minimize the density stipes observed after long TOF times. It is to be noted that this compensation is not perfect (estimated residual $\Delta U < k_B\times 0.2~$nK) and could be partially responsible for seeding the azimuthal modulational instability.

\subsection{Deterministic vortex creation and imaging}
\label{App:VCr}

We use a dynamic repulsive potential patterned by the DMD to deterministically create vortices in the 2D superfluids. Our procedure comprises of a repulsive circular beam of an initial radius of $\approx4~\mu$m and a potential height $\approx k_B\times 47~$nK (optimized for our sample density) slowly sweeping through the superfluid while splitting into two beams of the same diameter at a 65$^\circ$ separation angle. The beams move approximately $12~\mu$m in 400~ms, corresponding to a linear speed of $\approx 0.04c $, where $c=\hbar\sqrt{n_ig_i}/m\approx 0.76~$mm/s is the bulk sound speed. 
 
Ideally, this procedure creates a density defect with a 2$\pi$-phase winding around each beam center but with opposite circularity, forming a vortex dipole. 
We arrange the repulsive potential sweep such that one beam ends up at the center of the superfluid while the other one stops near the box boundary~\cite{gertjerenken2016generating}, as illustrated in Fig.~1(a). After each beam reaches its final position, the beam size is slowly shrunk down to zero in another 490~ms. From single-shot absorption images, we observe a single vortex at the superfluid center with nearly 90\% probability. Near the edge of the 2D gas, we observe the second vortex for less than 10\% of the experimental repetitions.

While we have performed GPE simulations using the chopstick method~\cite{samson2016deterministic,gertjerenken2016generating} as depicted in Sec.\ref{App:GPE}, we have experimentally searched for the proper conditions for deterministic vortex creation and pinning, which largely depends on the potential height and the sweep speed of the chopsticks, as well as the time over which we ramp off the size of the potential. The sweep speed and the ramp-off time (490 ms) are optimized so that the vortices are pinned to the chopstick beam centers with high probability. We observed that a faster sweep generates more than one randomly positioned density defect. Similarly, a faster ramp-off of the chopstick beam size either leads to a lower vortex creation probability or results in vortex unpinning. We have also optimized the size of the chopstick beams and the angle of divergence, with respect to the sample size, to reach higher vortex generation probability. Detailed systematic studies of our deterministic 2D vortex creation will be presented elsewhere.

\begin{figure}[h]
\centering
\includegraphics[width=0.4\columnwidth]{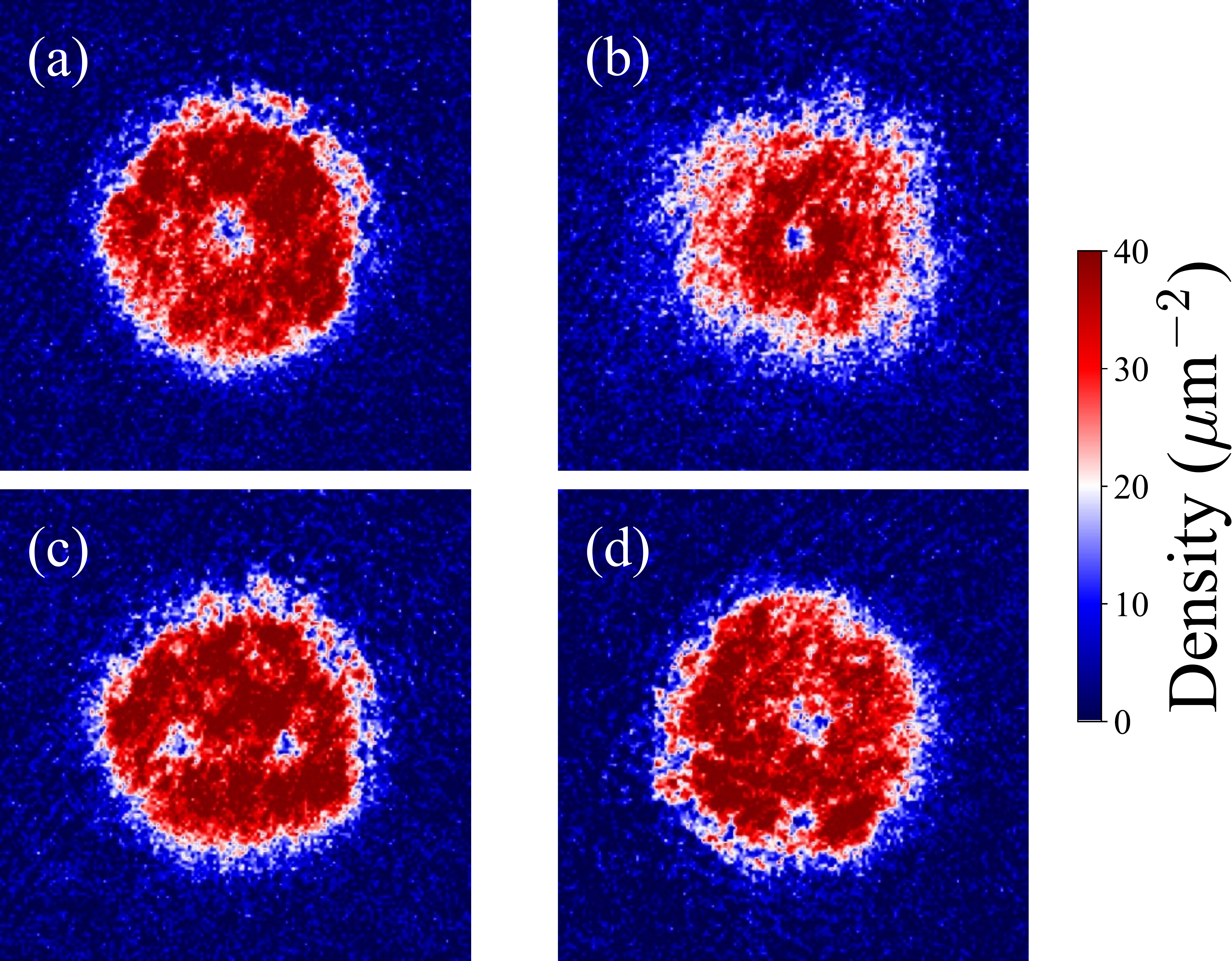}
\caption{Single-shot images of (a) a single vortex imaged in-situ after an interaction ramp from $g\approx0.1$ to $0.0056$, (b) a single vortex imaged after a 40~ms 2D TOF at $g\approx0.0056$, (c) and (d) vortex dipoles imaged in-situ.}\label{figSM:VCr}
\end{figure}

Figure~\ref{figSM:VCr} illustrates sample images of vortices created in 2D superfluids. Figure~\ref{figSM:VCr}(a) shows a single vortex imaged in-situ in the repulsive interaction regime. In order to clearly resolve the vortex core, we slowly ramp the interaction from $g\approx0.1$ down to $0.0056$ in 40~ms to enlarge the core size before imaging. Figure~\ref{figSM:VCr}(b) presents another image of a vortex prepared under the same interaction ramp procedure but with an additional 40~ms of 2D TOF time at the same interaction $g\approx0.0056$. The vortices remain stable under a variety of 2D TOF conditions, including long TOF under a strong repulsive interaction. To clearly image the vortex core, however, we always need to ramp the interaction down to a weaker value prior to imaging. 

Figure~\ref{figSM:VCr}(c) shows an in situ image of a vortex dipole created symmetrically about the trap center. Figure~\ref{figSM:VCr}(d) depicts another image of a vortex dipole of a different orientation, with the second vortex created near the edge of the superfluid at the lower side of the image. This is the configuration used for the main experiment. The probability of observing a second vortex near the edge, as shown in (d), is $\lesssim 10\%$. Although not demonstrated in this study, bringing the second chopstick beam out of the box can in principle completely anihilate the second vortex~\cite{gertjerenken2016generating}.

 \subsection{Formation of Townes solitons}\label{App:TS}
\begin{figure}[t]
\includegraphics[width=0.4\columnwidth]{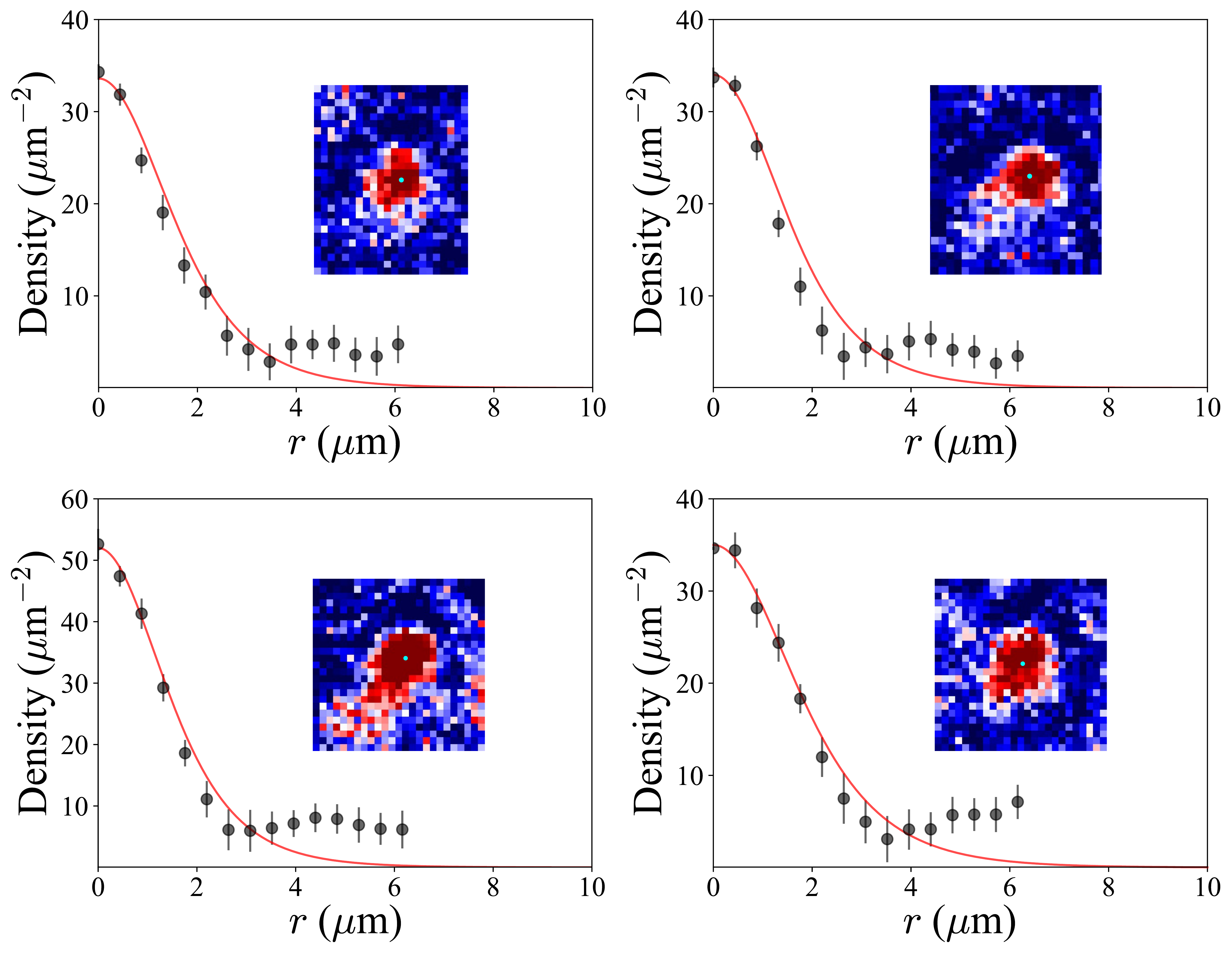}
\caption{Comparison of density blobs with the density profile of Townes solitons. Each panel shows a radially averaged density profile (filled circles) from the density image shown in the inset. The images in the top row are randomly selected from samples of interaction strength $g_f\approx-0.011$ and TOF time $\tau=49~\mathrm{ms}$, while the images in the bottom row are from samples of $g_f\approx-0.0077$ and $\tau=53~\mathrm{ms}$. Cyan dots in the images mark the center used for radial averaging. Solid lines are fits using Eq.~(\ref{Eq:TS}). }\label{figSM:TS}
\end{figure}

 Figure~1(c) shows that, at long enough TOF times, the gas fragments into a ring of disjoint density blobs. To test whether these blobs are Townes solitons, we numerically solve the scale-invariant stationary GPE, 
 \begin{equation}
-\frac{1}{2}\left(\frac{\partial^2 }{\partial R^2}+\frac{1}{R} \frac{\partial }{\partial R} -\frac{s^{2} }{R^2}\right)\phi-|\phi|^2 \phi=\tilde{\mu} \phi \,,
\end{equation}
 
 with zero winding number $s=0$ to obtain the steady-state solution $\phi(R)=\phi_{TS}(R)$. Here, $R=\sqrt{n_p\left|g_f\right|}r$ is the rescaled radial coordinate, $n_p$ is the peak density, and $g_f<0$ is the attractive interaction strength. The density profile of Townes solitons must have the form~\cite{chiao1964self,chen2021observation}
	\begin{equation}
		n(r)=n_p\left|\phi_{{TS}}\left(\sqrt{n_p\left|g_f\right|} r\right)\right|^2\,.\label{Eq:TS}
	\end{equation}
In Fig.~\ref{figSM:TS}, we randomly pick some isolated and round blobs to fit their radially averaged density profiles with Eq.~(\ref{Eq:TS}), using the peak density $n_p$ as the only free parameter.  
Indeed, we find reasonable agreement between the density profiles and those of Townes solitons except near the tails at $r\gtrsim 4~\mu$m. The deviation may be attributed to collapsed remnant gas in the background, dispersed blobs that failed to form Townes solitons, or the close proximity with other waves. 

\subsection{Extended mean-field analyses of radial wave collapse and the azimuthal modulational instability}\label{App:GPE}
We numerially evaluate the vortex collapse dynamics via the time-dependent GPE emulating the experimental procedures, including vortex generation via the chopstick potential and the interaction ramp-down followed by the 2D TOF. We begin with the time-dependent 2D GPE (without the three-body loss term),
\begin{align}
i\hbar \frac{\partial}{\partial t}\psi(\bv{r}) = \bigg(-\frac{\hbar^2}{2 m}\boldsymbol{\nabla}^2 + W(\bv{r}) + g|\psi(\bv{r})|^2+
 V_{c}(\bv{r})\bigg) \psi(\bv{r}) \,,
\label{impurityGPE}
\end{align}
where $\psi(\bv{r})$ is the condensate wavefunction, $W(\bv{r}) = W_0 [1+\operatorname{erf}((r-r_t)/\sigma_t)]/2$ denotes the circular confining potential,
$W_0 = k_B \times 47$ nK is the potential height, $r_t = 26$~$\mu$m is the radius, and $\sigma_t = 4.8$ $\mu$m the width of the confining potential. Here, $g = g_{3D}/(\sqrt{2\pi}\l_z)$ refers to the effective 2D interaction strength obtained by considering a tight confinement along the $z$-direction \cite{Tiwari:tracking}, where $g_{3D} = 4\pi\hbar^2a_s/m$, $a_s = 100 a_0$, and $a_0$ is the Bohr radius. 
The last term in \eref{impurityGPE} is the chopstick potential $V_c(\bv{r}) = V_0 [1-\operatorname{erf}((|\bv{r}-\bv{r}_0|-r_c)/\sigma_c)]/2$, which is initially centered around $\bv{r}_0= (x_0,y_0)=(-10.0~\mu$m, $-7.0~\mu$m) with $V_{0}=W_0$, $r_c = 3.5$ $\mu$m the radius, and $\sigma_c = 3.0$~$\mu$m. 

Using a high-level computing language XMDS \cite{xmds:docu,xmds:paper}, which invokes an adaptive step-size 8/9’th order Runge-Kutta method for time-stepping, and the Fast-Fourier-Transform for calculating the Laplacian, we numerically solve \eref{impurityGPE} in the imaginary time evolution and obtain the ground state with the surface density $n_i = |\psi|^2 \approx 25$~$\mu$m$^{-2}$ in the bulk. The density falls to half of this value at the radial distance $r_i \approx 21~\mu$m as in the experiment. After finding the ground state, we move and split the chopstick potentials at a separation angle $65^{\circ}$ to their final position within $400$~ms while simultaneously solving the GPE in real-time. The final position of the first chopstick is $(x_{f},y_{f}) = (0,0)$, whereas we move the second chopstick out of the box. The size of the chopsticks is reduced to zero at their final positions within an additional 500~ms, which results in only a single vortex at the center of the wavefunction. 

Following the vortex preparation sequence, we modify the mean-field wavefunction $\psi(\bv{r})$ by seeding noise in the form of the Bogoliubov modes, 
\begin{align}
\eta(\bv{r}) = \psi(\bv{r}) + \frac{1}{\sqrt{2}}\sum_{\bv{k}}  [\alpha_\bv{k} u_k(\bv{r}) - \alpha^*_\bv{k} v^*_k(\bv{r}) ],
\label{Twavefunction}
\end{align}
where $\alpha_\bv{k}$ is a random complex Gaussian noise satisfying the relation $\overline{\alpha_\bv{k}\alpha_\bv{k'}} = 0$, $\overline{\alpha_\bv{k}\alpha^*_\bv{k'}} = \delta_{\bv{k}\bv{k'}}$, and $\overline{\cdots}$ is a stochastic average. Here, $u_{k}(\bv{r})$ and $v_{k}(\bv{r})$ are the Bogoliubov modes assuming a homogeneous 2D condensate density. This procedure corresponds to seeding on-average 0.5 phonons per $\bv{k}$-mode in the modified wavefunction, simulating fluctuations at zero temperature~\cite{blakie2008dynamics}.

We then perform the GPE calculations in which we phenomenologically incorporate a three-body loss term $-i \hbar \zeta K_{3}|\psi(\bv{r})|^4\psi(\bv{r})/2$ in ~\eref{impurityGPE} with a constant loss rate $K_3 = 1.2\times10^{-26}$cm$^{6}/$s. This rate is around 10 times larger than a previously measured value using thermal gases prepared at $a_s\approx 0$~\cite{kraemer2006evidence}. The factor $\zeta = 1/(6\sqrt{3}\pi l_z^2)$ takes into account Bose symmetrization and integration along the $z$-axis. We evolve the GPE for 6~ms by inserting the wavefunction of Eq.~\bref{Twavefunction} and then slowly ramp down the trap potential height $W_0$ to zero within $3$~ms while simultaneously ramping the interaction $g$ to an attractive value $g_f<0$ within $5$~ms as mentioned in the main text.  
We then continue to evolve the GPE for additional time $\tau$ at each final interaction strength. Ten different initial random noise samples are considered for each time evolution.

\begin{figure}[t]
\includegraphics[width=0.6\columnwidth]{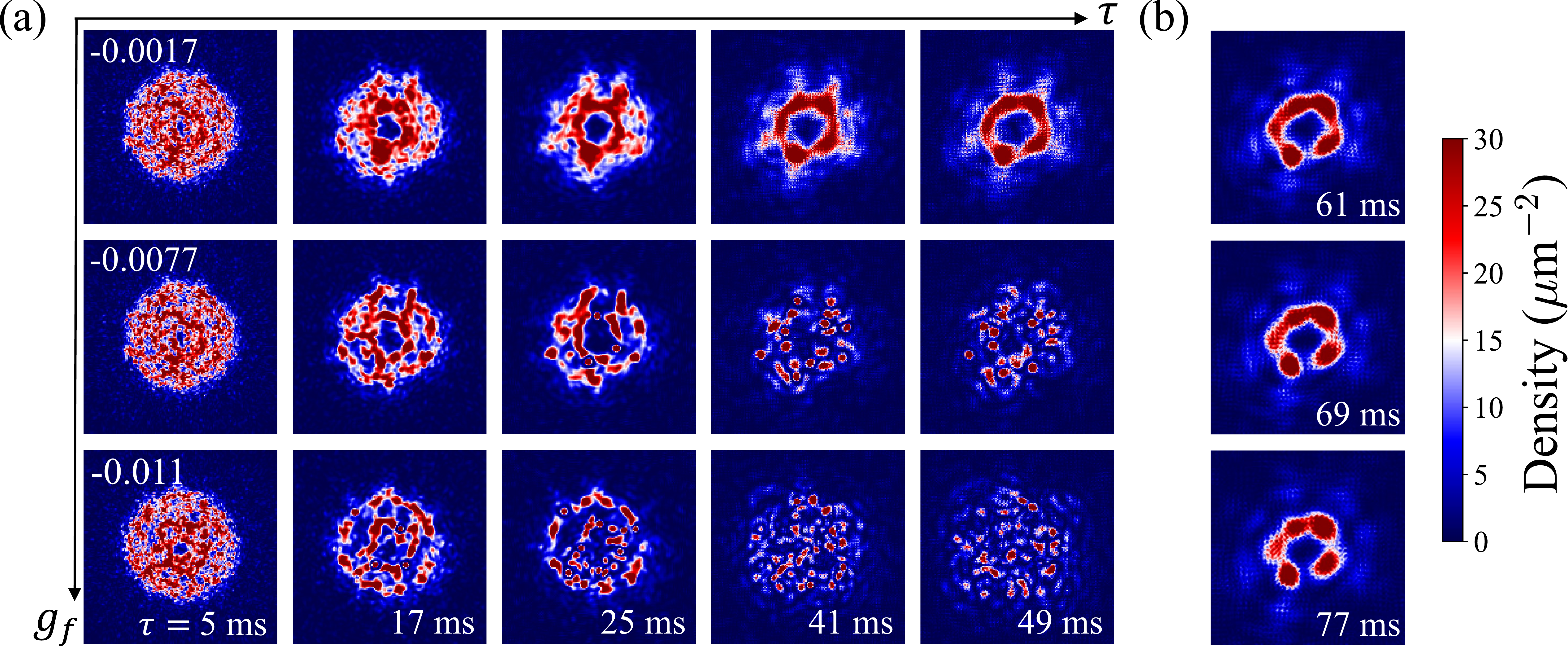}
\caption{\label{FigSM:densEvol} Dynamics of the density evolution, $n(\bv{r},\tau) = |\eta(\bv{r},\tau)|^2$, obtained from the GPE calculation with (a) a variable interaction strength $g_f$ and different hold time $\tau$, and (b) $g_f=-0.0017$ at longer times as in Fig.~1(c, d). Image size: (75 $\mu$m)$^{2}$.
}
\end{figure}

\begin{figure}[t]
\includegraphics[width=0.5\columnwidth]{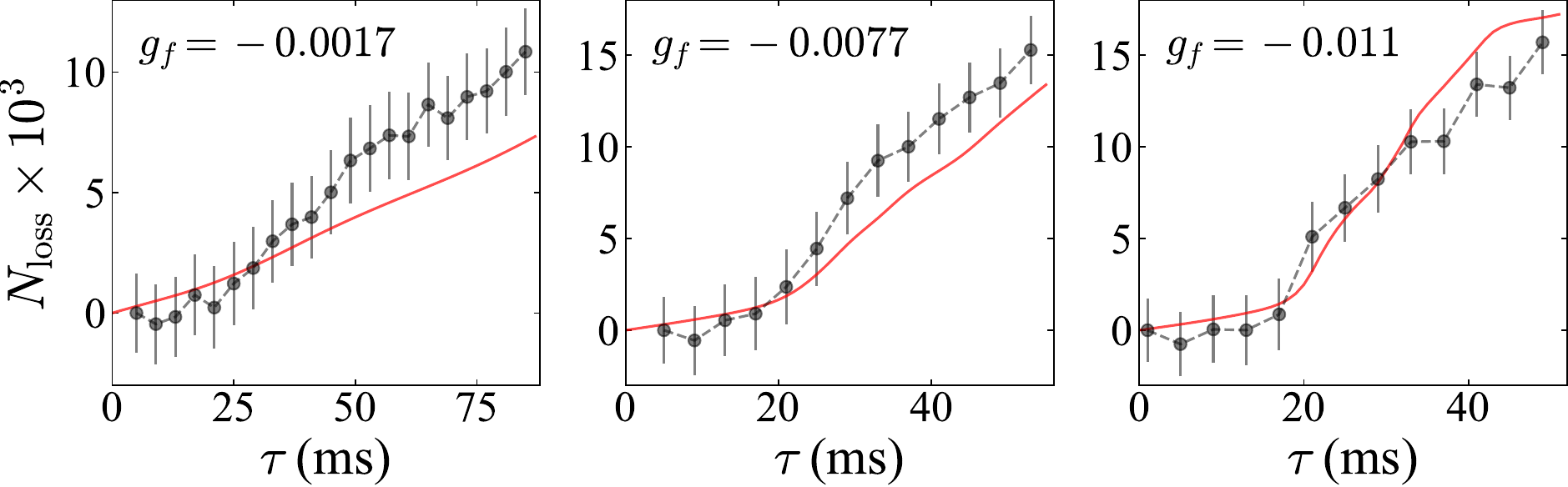}
\caption{\label{Loss}  Atom loss for three different values of $g_f$. The red solid lines are obtained from the simulations, and the black circles are from the experiment. 
}
\end{figure}

Figure~\ref{FigSM:densEvol} shows the evolution of the density $n(\bv{r},\tau) = |\eta(\bv{r},\tau)|^2$, in a single realization, for three different interaction strengths. The simulation employs $256\times 256$ spatial grid points for a box of $64$~$\mu$m in length. The first row of panels in \fref{FigSM:densEvol}(a), and \fref{FigSM:densEvol}(b), are snapshots of the density for $g_f\approx -0.0017$ ($a_s = -1.65a_0$). Enlargement of the density defect surrounded by an enhanced ring structure similar to the experiment can be clearly seen (see Fig.~1). Wave fragmentation and enhanced density blob features appear to be more prominent as the interaction strength becomes more attractive, as shown in the last two rows of panels in \fref{FigSM:densEvol}(a). Inclusion of the three-body loss term suggests delay of the wavefunction collapse and soliton patterns are observed in the final snapshots. The corresponding atom loss for these three interaction strengths are presented in \fref{Loss} together with atom loss measured in the experiments. One can see excellent match between the simulation and experimental data for higher interaction strengths. A slightly larger atom loss occurs in the experiment for lower interaction strengths. We attribute this to imperfect parameter matching for different experimental conditions such as slightly higher initial density or deviation of the three-body loss coefficients. Notice that within the GPE simulations, we have assumed that all parameters are fixed except for $g_f$. 

In \fref{fig2_radial}, we numerically evaluate the time evolution of the radial density profiles averaged over ten realizations of initial random noise. To identify the time scales for radial convergence in the density profiles, we adopt the procedure for evaluating the variance $V$ as described in the main text.

\begin{figure}[t]
\includegraphics[width=0.6\columnwidth]{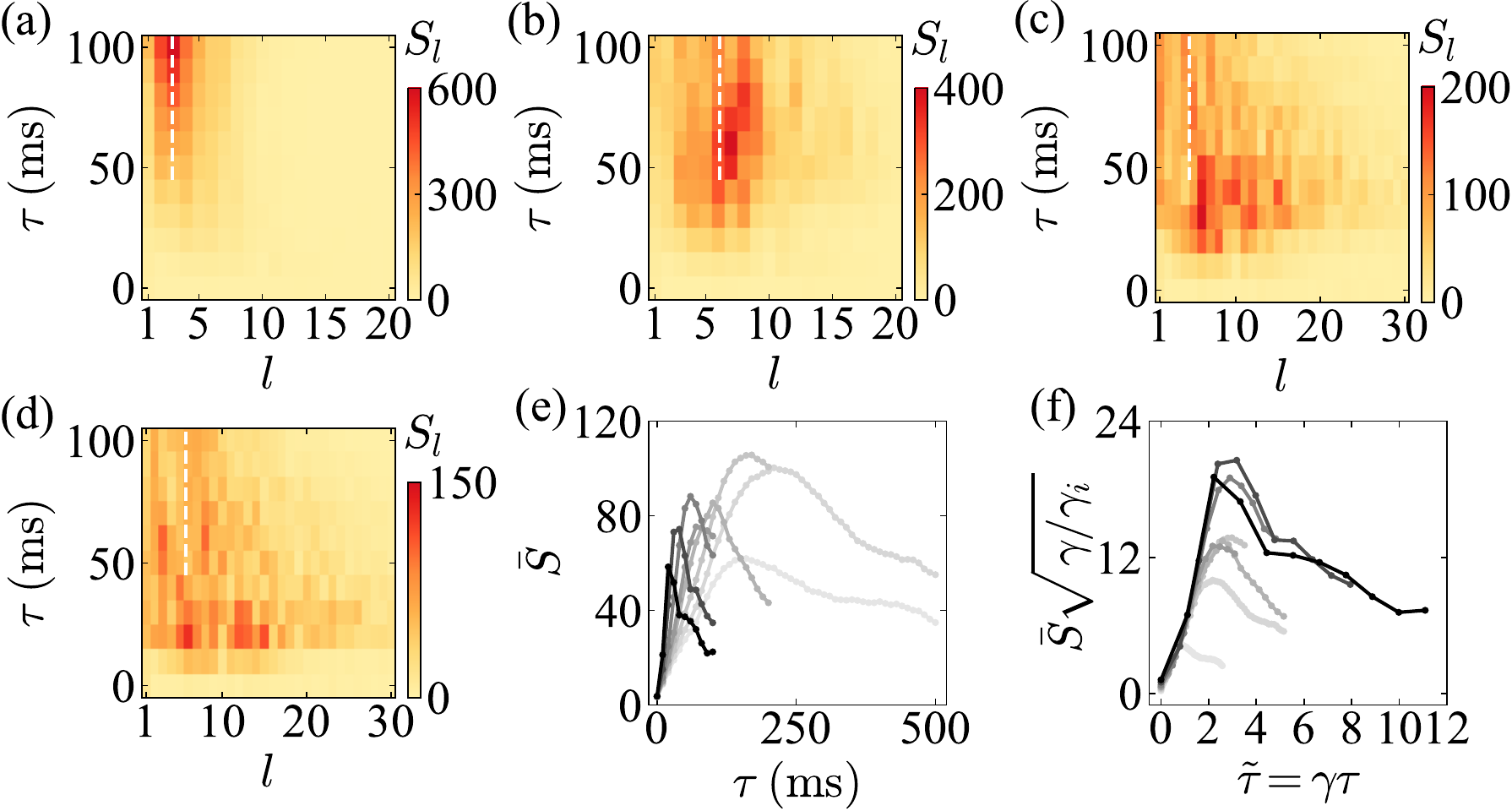}
\caption{Angular power spectra $S_l$ evaluated from the time-dependet GPE with $g_f\approx$ (a) -0.0017, (b) -0.0047, (c) -0.0077, and (d) -0.011, respectively. White dashed lines mark the peak $l_p$. (e) Averaged power spectra $\bar{S}$ versus time and (f) rescaled power spectra versus rescaled time obtained at the interaction strengths $g_f\approx$ -0.011, -0.0077, -0.0047, -0.003, -0.0025, -0.0017, -0.001, and -0.0005 (solid curves, from dark to light gray), respectively.}\label{fig:GPEspectra}
\end{figure}

To analyze the azimuthal modulational instability and compare with the experimental data, we evaluate the angular power spectra as shown in \fref{fig:GPEspectra}, from which we identify the peak position of the power spectrum $S_l$ as described in Sec.\ref{App:peak}. The averaged power spectrum $\bar{S}$ is evaluated within the same range ($l=1\sim 30$) described in the main text. The results are shown in \fref{fig:GPEspectra}(e). Rescaled spectra $(\gamma/\gamma_i)^\alpha\bar{S}$ versus rescaled time $\tilde{\tau} = \gamma\tau$ are shown in \fref{fig:GPEspectra}(f) with an empirical scaling exponent $\alpha=0.5$. The rescaled spectra are compared with the experimental data in Fig.~3(e), both exhibiting a remarkable universal behavior before the fragmentation time $\tilde{\tau}\lesssim 2$. We attribute the nonuniversal curves at $|g_f|\lesssim 0.001$ to finite size effects when $r_i$ approaches $\xi$. 

\subsection{Determination of the peak position in the power spectrum $S_l$}\label{App:peak}
\begin{figure}[h]
\includegraphics[width=0.5\columnwidth]{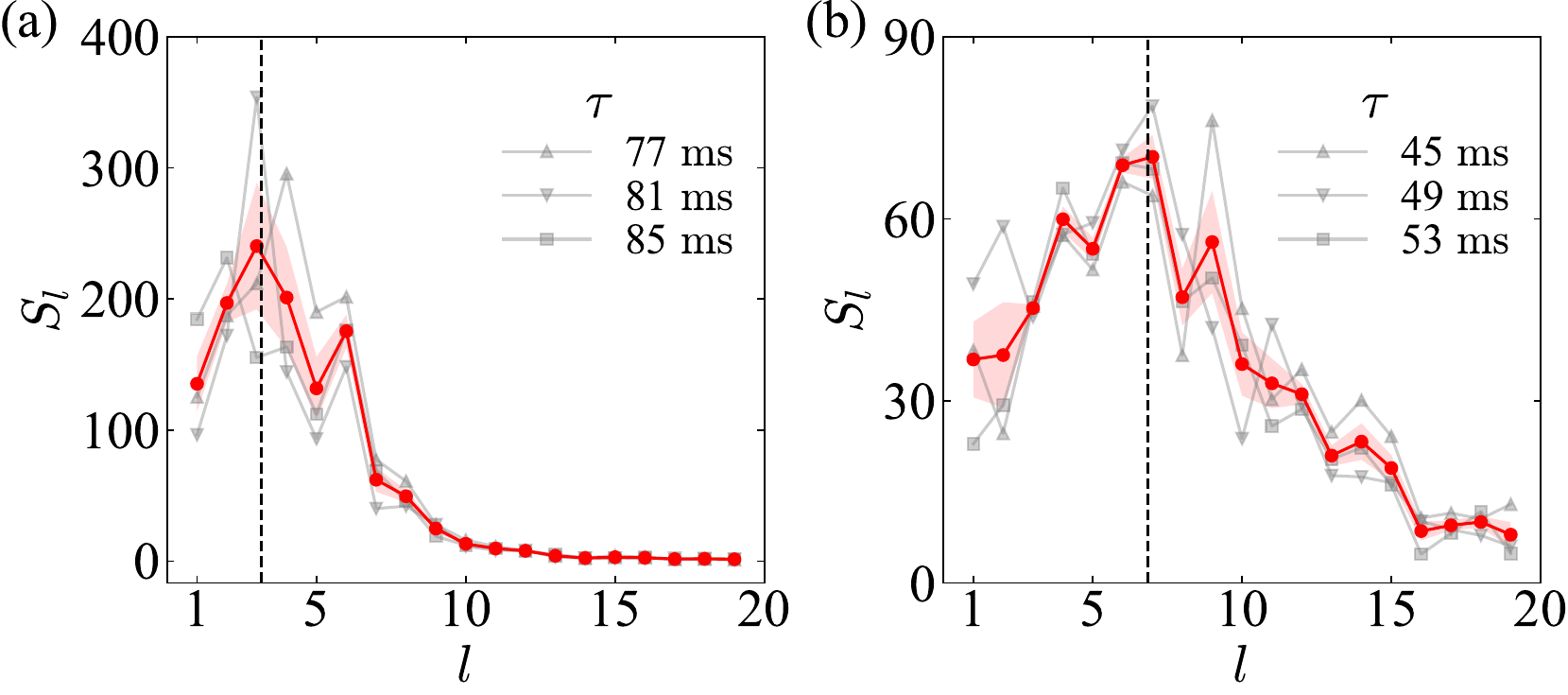}
\caption{Angular power spectrum $S_l$ experimentally measured at three longest TOF times with (a) $g_f\approx-0.0017$ and (b) $g_f\approx-0.0077$, respectively. Red circles represent the mean of the power spectra and the shaded band indicates the corresponding standard deviation of the mean. Black dashed line indicates $l_p$ determined from the mean of individually identified peak-$l$ in each spectrum.}\label{fig:peak}
\end{figure}

In experiments and in numerical simulations, the angular power spectrum $S_l$ developes a distinct peak at each interaction $g_f$. To determine the most unstable angular mode, we identify the mode number $l$ of the maximum $S_l$ measured at seven longest TOF times for each $g_f$ (or at times $\tau\leq 100~$ms for GPE simulations) and then compute their mean value $l_p$ and the standard error. The results are provided in Fig.~3(c). In Fig.~\ref{fig:peak}, we compare $l_p$ (black dashed line) with the power spectra measured experimentally for three TOF times. 

\subsection{Variational analysis of the azimuthal modulational instability}\label{App:AMI}
When the interaction strength of a homogeneous, non-rotating 2D superfluid is quenched to a negative value, a modulational instability will manifest, with self-amplifying density waves to fragment a sample into many density blobs~\cite{chen2020observation}. The most unstable wavenumber is $k_{\mathrm{MI}}=\sqrt{2n_i|g_f|}$, where $n_i$ is the initial atomic density and $g_f$ is the attractive interaction strength. For a circular sample carrying one quantum vortex at the center, the instability may be analyzed in discrete angular modes. However, for a sample whose initial radius is several times larger than the characteristic interaction length $\xi = \pi /k_{\mathrm{MI}}$, one may still expect a local manifestation of the modulational instability, as already hinted in the multi-ring fragmentation observed in Fig.~1(c). If the instability shares the same length scale as the modulational instability, the most unstable angular mode can be expressed as $l_{\mathrm{MI}}= \bar{r} k_{\mathrm{MI}}=\bar{r} \sqrt{2n_i|g_f|}$, where $\bar{r}$ is a mean radial position of the sample. 

Our intuitive picture can be supported by a variational analysis detailed in~\cite{caplan2009azimuthal,caplan2012existence}, which assumes a wavefunction of the form $\psi(r,\theta)=\phi(r)A(\theta,t)$ with a fixed radial function $\phi(r)=\sqrt{n(r)}$ and derives the nonlinear Schr\"{o}dinger equation for $A(\theta,t)$ using a variational method. By assuming a vortex state of $A(\theta,t) = e^{i(\pm\theta + \mu t)}$ with small perturbations in discrete $l$ angular modes and performing analyses for the azimuthal modulational instability, one finds the eigenfrequencies of the $l$-th mode to be 
\begin{equation}
\omega_l = \pm \gamma_\mathrm{va} \sqrt{l^2(l^2-2 l_p^2)}\, , \label{EqSM:freq}
\end{equation}
where $\gamma_\mathrm{va} = \frac{\hbar}{m}\frac{\int \frac{n(r)}{r^2} rdr}{\int n(r) rdr}$, $l_p = \sqrt{2|g_f|\mathcal{C}}$, and $\mathcal{C} = \frac{\int n(r)^2 rdr}{\int \frac{n(r)}{r^2} rdr}$. According to Eq.~\eqref{EqSM:freq}, $\omega_l$ becomes purely imaginary when $l<\sqrt{2}l_p$ and the modes are unstable. The nearest integer from $l_p$ is the most unstable mode with the largest imaginary frequency. We may carry out the radial integrations by using experimentally measured radial density profiles for $n(r)$ in the integrant. We may also link the most unstable mode $l_p$ with the most unstable wavenumber $k_\mathrm{MI}$ in the modulational instability as
\begin{equation}
    l_p = \bar{r}_v\sqrt{2|g_f| n_i} = \bar{r}_v k_\mathrm{MI} \, ,
\end{equation}
and define the effective radius for the azimuthal modulational instability as $\bar{r}_v \equiv \sqrt{\mathcal{C}/n_i}$. While the bove variational analysis is most accurate with a fixed radial density profile, in Fig.~3(c) we estimate $l_p$ using $\bar{r}=\bar{r}_v$ from the density profile at the saturation time $\tau_\theta$ of $\bar{S}$ when the azimuthal fragmentation takes place. 

\subsection{Azimuthal modulational instability of a vortex soliton}\label{App:AMI_VS}
\begin{figure}[t]
\includegraphics[width=0.3\columnwidth]{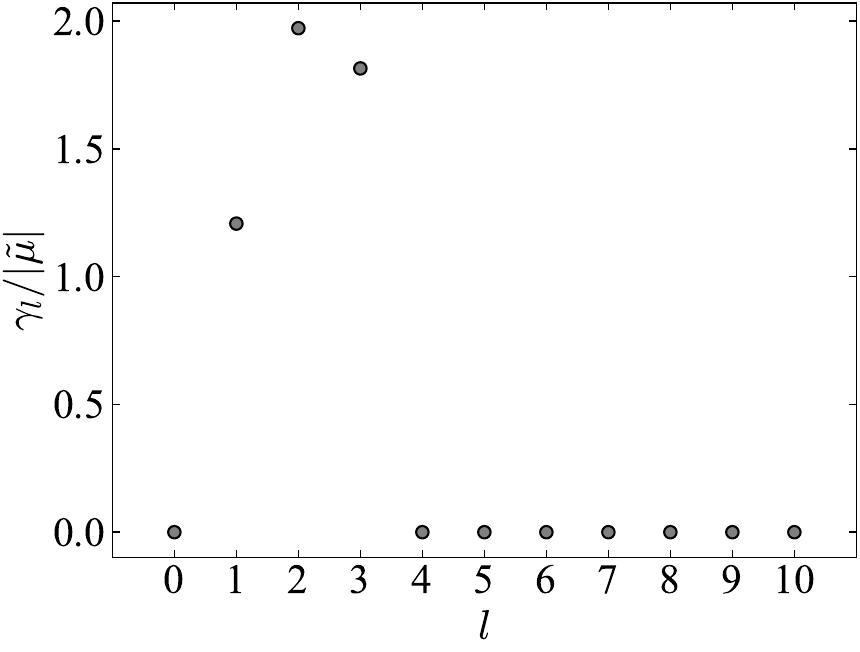}
\caption{Azimuthal modulational instability of a vortex soliton. $\gamma_l=|\mathrm{Im}(\omega_l)|$ denotes the growth rate at the $l$-th mode. It is only non-zero for $ l=1,2,3 $, and the most unstable mode is $ l=2$.}\label{fig:mode}
\end{figure}

The azimuthal modulational instability of a vortex soliton can be directly analyzed in a Bogoliubov analysis~\cite{edwards1996zero,saito2002split}. We do this by introducing small perturbations to the scale-invariant stationary-state solution of a vortex soliton,

\begin{equation}\label{modes}
    \psi(R,\theta,\tilde\tau)=\left\{\phi_{\mathrm{vs}}(R)+\sum_l\left[ u_l(R)e^{-i\omega_l\tilde\tau+il\theta}+v_l^*(R)e^{i\omega_l^*\tilde\tau-il\theta}\right]\right\} e^{-i\tilde\mu \tilde\tau\pm i\theta},
\end{equation}

where $\phi_{\mathrm{vs}}(R)$ is the radial wavefunction, $R=\sqrt{n_p\left|g_f\right|}r$ is the rescaled radial coordinate, $\tilde\tau=\gamma_p\tau$ is the rescaled time with $\gamma_p=\hbar n_p |g_f|/m$, and $\tilde{\mu}=\mu/(\hbar\gamma_p)$ is the scaled chemical potential as in the main text. Here, $u_l$ and $v_l^*$ are the Bogoliubov coefficients for the $l$-th angular mode, and $\omega_l$ is the corresponding eigenvalue. If the imaginary part of $\omega_l$ is non-zero, the mode becomes unstable with an exponential growth rate $\gamma_l=|\mathrm{Im}(\omega_l)|$. Substituting Eq.~\eqref{modes} into the scale-invariant 2D GPE and gathering all the terms evolving in the form of $e^{-i\omega_l\tilde\tau}$ and $e^{i\omega_l^*\tilde\tau}$, we obtain the Bogoliubov equations in the matrix form

\begin{equation}
    \left(\begin{matrix}
        -\frac{1}{2}\left[\frac{\partial^2}{\partial R^2}+\frac1R\frac{\partial }{\partial R}-\frac{(s+l)^2}{R^2}\right]-(2|\phi_{\mathrm{vs}}|^2+\tilde\mu)&-\phi_{\mathrm{vs}}^2\\\phi_{\mathrm{vs}}^{*2}&\frac{1}{2}\left[\frac{\partial^2}{\partial R^2}+\frac1R\frac{\partial }{\partial R}-\frac{(s-l)^2}{R^2}\right]+(2|\phi_{\mathrm{vs}}|^2+\tilde\mu)
        
    \end{matrix}
    \right)
    \left(\begin{matrix}
        u_l\\v_l
    \end{matrix}
    \right)=\omega_l
    \left(\begin{matrix}
        u_l\\v_l
    \end{matrix}
    \right)\,,
\end{equation}

and further discretize the coordinate $R$ and wavefunctions $(u_{l}, v_{l}, \phi_{\mathrm{vs}})$ with 500 grid points of step size $\Delta R=0.03$. Here, $\left(\frac{\partial^2}{\partial R^2}\right)_{ij}=(\delta_{i,j-1}-2\delta_{i,j}+\delta_{i,j+1})/\Delta R^2$ and $\left(\frac1R\frac{\partial }{\partial R}\right)_{ij}=(-\delta_{i,j-1}+\delta_{i,j+1})/(2\Delta R R_i)$, where $i(j)$ is the grid index and $\delta_{i,j}$ is the Kronecker delta. We calculate the growth rate $\gamma_l$ by solving the above eigenvalue problem. The result is shown in Fig.~\ref{fig:mode}, where the most unstable mode is at $l=2$, corresponding to the blue dashed line in Fig. 3(c). Our result is consistent with the split instability discussed in Refs.~\cite{saito2002split,mihalache2006vortex}.

\subsection{Discrepancy in the angular power spectrum between experiment and simulation}\label{App:APS_sim_exp}
We observe that the scaled power spectra of our GPE simulations are approximately a factor of 2 higher than the experimentally obtained results in the early-time growth dynamics. The GPE spectra peak at even higher amplitudes followed by a monotonic decay after fragmentation. During the later time dynamics of fragmentation and collapse, we suspect that microscopic details such as inevitable three-body recombination mechanisms may be strong enough, altering the self-similar behavior. Therefore, we focus on reconciling the discrepancy in the growth phase, where atom loss is less relevant.  Correspondingly, we consider additional systematic effects within the GPE analysis. Specifically, we have explored the following plausible explanations: (1) We have tested the effect of non-smoothly turning off the chopstick potentials, which could introduce additional density perturbations during the vortex generation process. (2) Additionally, we have introduced small potential corrugations in our simulations which could influence the dynamics of azimuthal modulational instability. (3) We have further performed extensive GPE simulations, utilizing samples containing up to 100 realizations, in order to ensure convergence of the calculated angular power spectrum. (4) We have convolved the simulated density profiles with the point spread function of our imaging system to account for the finite resolution effect. (5) We have carefully calibrated the density conversion in absorption imaging and have incorporated the effect of nonlinear density conversion at high densities beyond $150/\mathrm{\mu}$m$^{2}$ when evaluating the numerical power spectra.

\begin{figure}[h]
\includegraphics[width=0.8\columnwidth]{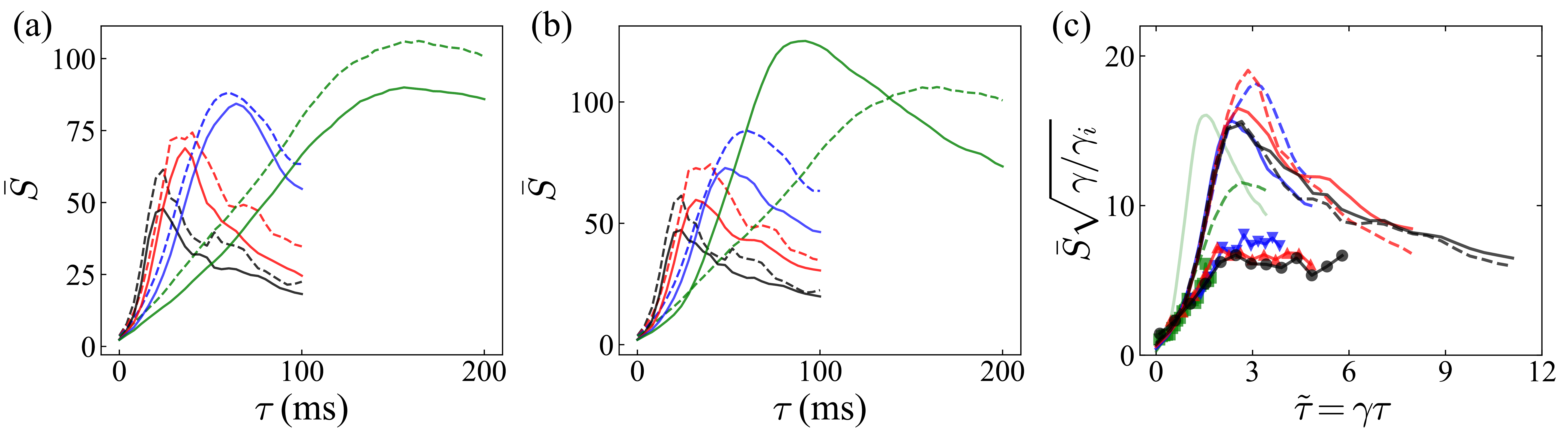}
\caption{(a,b) Comparison of the averaged power spectra $\bar{S}$ as shown in Figs.~3 and 4 (dashed lines, color scheme: $g_{f}\approx -0.011$ (black), $-0.0077$ (red), $-0.0047$ (blue), and $-0.0017$ (green)) with the calculated spectra (solid lines, same colors) after we incorporate the effects discussed in Sec.\ref{App:APS_sim_exp} with (a) no potential corrugation $\Delta U=0$ and (b) a corrugation depth $\Delta U=k_B\times0.3$ nK. 
(c) Scaled power spectra versus rescaled time for the new simulations in (a) and (b), with (solid lines) and without (dashed lines) the corrugated potential. Colored symbols show the experimental data as in Fig. 4(a). } 
\label{fig:discrepancy}

\end{figure}

All the modifications discussed above, however, cannot fully resolve the discrepancy in the magnitude of the rescaled angular power spectra. Incorporating all the effects, except for a potential corrugation, cumulatively lowers the peak value of the averaged angular power spectrum by around 22\%, 7\%, 4\% and 15\% for $|g_f|\approx 0.011$, $0.0077, 0.0047$ and $0.0017$, respectively. 
Adding a  potential corrugation $\Delta U=k_B\times 0.3$~nK additionally lowers the values to 23\%, 20\% and 18\% for the three more negative interactions. Figure~\ref{fig:discrepancy} displays a comparison between the power spectra derived from the GPE simulations discussed in Sec.\ref{App:GPE} and the modified results after incorporating all the effects discussed above (each averaged using 30 realizations). 
The self-similar scaling behavior remains robust and the rescaled spectra are nearly identical for $\tilde{\tau}\lesssim 2.5$. For $|g_f|\approx0.0017$ (green curves), the peak interaction energy ($\approx k_B\times0.19$ nK) is much lower than the intentionally exaggerated corrugation strength of $k_B\times0.3$ nK, leading to a sizable deviation from the rescaled universal curve in Fig. \ref{fig:discrepancy}(c) (green solid line). Adding all the effects reduces the self-similar GPE spectra to a value around 1.7 times higher than the experimentally measured spectra in the early-time growth. 

The remaining discrepancy may arise from other effects not considered in our GPE model, including those related to beyond mean-field physics. We note that the radial convergence of the samples to a vortex soliton-like profile is observed after averaging over $\sim 12$ shots. While the convergence is well captured by the mean-field GPE model, the azimuthal power spectra reflect the density fluctuations and correlations in single shots, which would be sensitive to effects of beyond mean-field physics. Hence, our analysis unveils the necessity to account for quantum fluctuation phenomena and also calls for future development of beyond mean-field theoretical methods, for instance, ranging from perturbative techniques~\cite{petrov2024beyond} to sophisticated ab-initio approaches~\cite{cao2017unified}.

Such highly unexplored developments are certainly worth pursuing in future theory endeavors benchmarked against experimental data but lie beyond the scope of the current investigation. Indeed, such comparisons would be non-trivial not only due to the fact that the aforementioned beyond mean-field techniques are numerically demanding in their own right but also because emergent quantum fluctuation phenomena have to be carefully distinguished in order to identify the ones that are responsible for the experimental observations. Finally, these explorations would pave the way for systematically probing the correlated character of the involved processes by relying on higher-order observables such as correlation functions and entanglement.

\end{document}